**COSMIC THERMOBIOLOGY**
*Thermal Constraints on the Origin and Evolution o f Life in the Universe*

**CHARLES H. LINEWEAVER[1] AND DAVID SCHWARTZMAN[2]**
[1]*School of Physics, University of New South Wales, Sydney, Australia*
[2]*Department of Biology, Howard University, Washington DC, USA*

**1. Thermal Determinism and the Narrow Roads of Prebiotic Chemistry**

Modern cosmology gives us the largest context within which to understand the emergence of life in the universe. Within this context, the transition from molecular to biological evolution is the most recent in a series of transitions that can be easily described as the result of decreasing temperature. In the big bang model for the evolution of the universe, as the universe expanded and cooled from arbitrarily high temperatures, a universal deterministic cascade of structure appeared. Here we describe how the expansion and decreasing temperature of the universe allowed atoms, molecules, stars, planets and life to form. We discuss the thermal constraints on terrestrial biogenesis and suggest that they apply to the earliest forms of life, not only on Earth, but anywhere in the universe.

Cosmology, physics and chemistry are deterministic in the sense that what we learn on Earth will be valid in the most distant parts of the universe. Biological evolution on the other hand seems to be dominated by quirky tinkering and historical contingency. We do not expect elephants or tuataras to have evolved anywhere else in the universe. If life emerged from non-life through a process of molecular evolution, there must be a continuum or a transition between these two paradigms. The first steps in the transition will be inevitable and temperature-induced. The next steps will be less deterministic and the final steps will be as highly contingent and unpredictable as life is today. Here we describe the first thermal constraints of an abiotic deterministic universe and the transition to a less-predictable biotic one. This thermal determinism is similar to the biochemical determinism of "Vital Dust" (deDuve, 1995). deDuve argues that biogenesis and much of what life is, is biochemically inevitable. Here we argue that biogenesis and much of what life is, is thermally inevitable.

**2. When Does the Temperature of the Universe Permit the Ingredients of Life to Exist?**

The very early universe was filled with radiation at arbitrarily high temperature (Fig. 1). At $\sim 10^{-33}$ seconds after the big bang, the universe cooled enough to allow matter and anti-matter to annihilate, converting rest mass into radiation while leaving a small excess (one part in one billion) of matter. This transition is known as baryogenesis and is the formation of all the matter in the universe. At $\sim 10^{-4}$ seconds after the big bang the thermal energy of the universe was low enough to allow the hot soup of quarks to condense – quarks fell together in triplets under the influence of the strong nuclear force and have remained stable ever since, as protons and neutrons. This epoch is known as the quark-hadron transition. At ~ 100 seconds after the big bang, nucleosynthesis began. The thermal energy of the universe decreased allowing protons and neutrons to bind together under the residual strong nuclear force to form the nuclei of the lightest elements: $^1$H, $^2$H, $^3$He, $^4$He, $^7$Li.

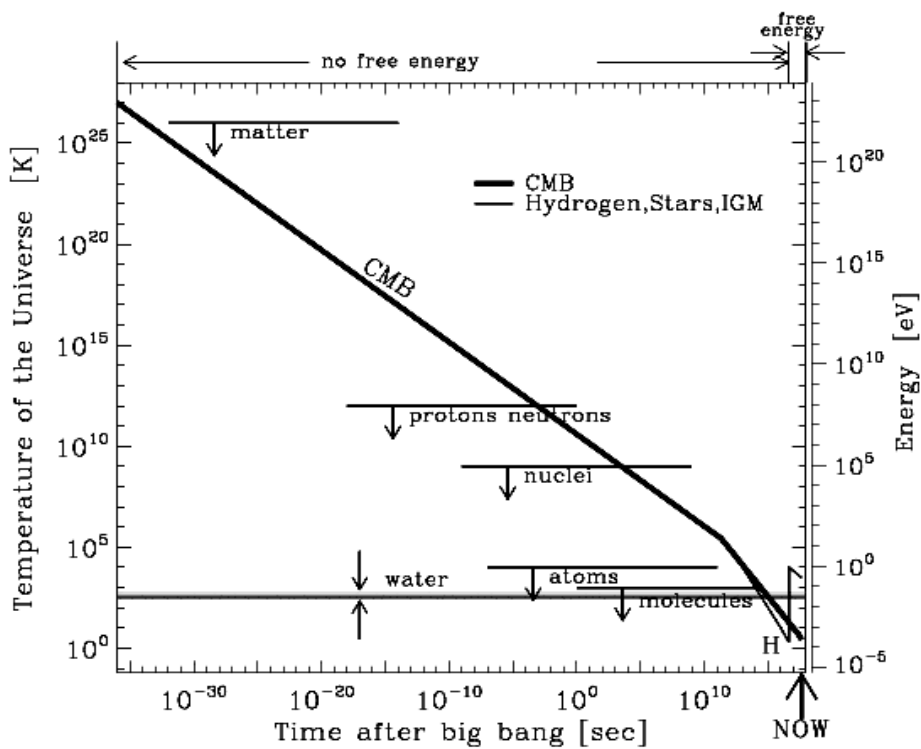

Figure 1. Transitions in the universe as the temperature decreases. Structures which freeze out as the universe cools include, matter, protons and neutrons, nuclei, atoms and molecules. See text and Fig. 2 for details.

For the first 400,000 years after the big bang all the matter was in thermal equilibrium with the sea of photons (cosmic microwave background: CMB) that filled the universe. Then the thermal energy of the CMB fell beneath the ionization energy of hydrogen. This allowed the electromagnetic force between electrons and protons to pull them together to form hydrogen atoms. Without charged particles to maintain thermal equilibrium, the temperature of the matter and the photons went their separate ways. The temperature of the photons cooled inversely proportional to the size of the universe, while the matter

TABLE 1. Thermal History of the Universe (see Fig. 1)
___________________________________________________________________

| Structure formed | Temperature [K] | Energy [eV] | Time after Big Bang [sec] |
| --- | --- | --- | --- |
| matter | $10^{15}$ | $10^{11}$ | $10^{-33}$ |
| protons/neutron | $10^{12}$ | $10^{8}$ | $10^{-4}$ |
| atomic nuclei | $10^{9}$ | $10^{5}$ | 100 |
| atoms | $10^{4}$ | ~1 | 400,000 years |
| molecules | $10^{3}$ | ~0.5 | few million years |
| aqueous macromolecules | $10^{2}$ | ~0.1 | 2 billion years |

(Kolb and Turner, 1990; Lewis, 1997)

cooled faster, inversely proportional to the square of the size of the universe. For the first time in the history of the universe two temperatures were needed to describe the universe. As the temperature of the hydrogen dropped below the dissociation energy of molecular hydrogen, hydrogen atoms in the densest parts of the universe paired up to form the first molecules in the universe: $H_2$. There were still no stars or planets. There was no $H_2O$ in the universe. Stars had not yet made oxygen, carbon or nitrogen. There was no life.

The universe continued to expand and cool. When clouds of hydrogen cooled beneath 100 K, the thermal energy decreased enough to allow the weakest force in the universe, gravity, to make the densest regions gravitationally collapse. Gravitational potential energy was converted to heat and lost to outer space. The clouds collapsed further, heated further and finally massive stars were formed. Their UV photons re-ionized the universe. Their supernova explosions induced the collapse of other dense regions. In Figs. 1 and 2 the thin line labeled "H" shows the temperature of the hydrogen. It shot up to 5,000 - 10,000 K as the first stars in the universe broke the equilibrium and provided free energy for the first time in the history of the universe (see region labeled "free energy" above the upper x axis).

About 1% of the normal material in the universe today is not hydrogen and helium. These heavy elements were not produced in the big bang. They were produced gradually by generations of stars. It is the presence of these heavy elements which allows small rocky planets to form when clouds of molecular hydrogen collapse to form stars and gaseous planets.

The Earth formed from a molten ball at ~2000 K about 4.56 Gyr ago (Allègre et al., 1995). The majority of the Earth's mass accreted from planetesimals within the first 100 million years of the Earth's formation (Halliday, 2000). With an initially molten surface, life could not have appeared. The transition from accretion to heavy bombardment included the formation of the Moon by the collision with a Mars-sized object ~4.5 Gyr ago (Hartmann and Davis, 1975; Halliday, 2001; Canup and Asphaug, 2001). We can infer from the dates and sizes of lunar impact craters, whose record goes back to when the Moon formed a solid crust (~4.44 Gyr ago, Sleep et al., 1989) that the surface of the Earth was periodically vaporized and covered with a 2000 K rock vapor atmosphere which lasted for several thousand years (Hartmann et al., 2000; Sleep et al., 2001). These conditions were probably an effective and recurring autoclave for sterilizing the earliest life forms or more generally frustrating the evolution of life. A steadily decreasing heavy bombardment continued until ~ 3.8 Gyr ago.

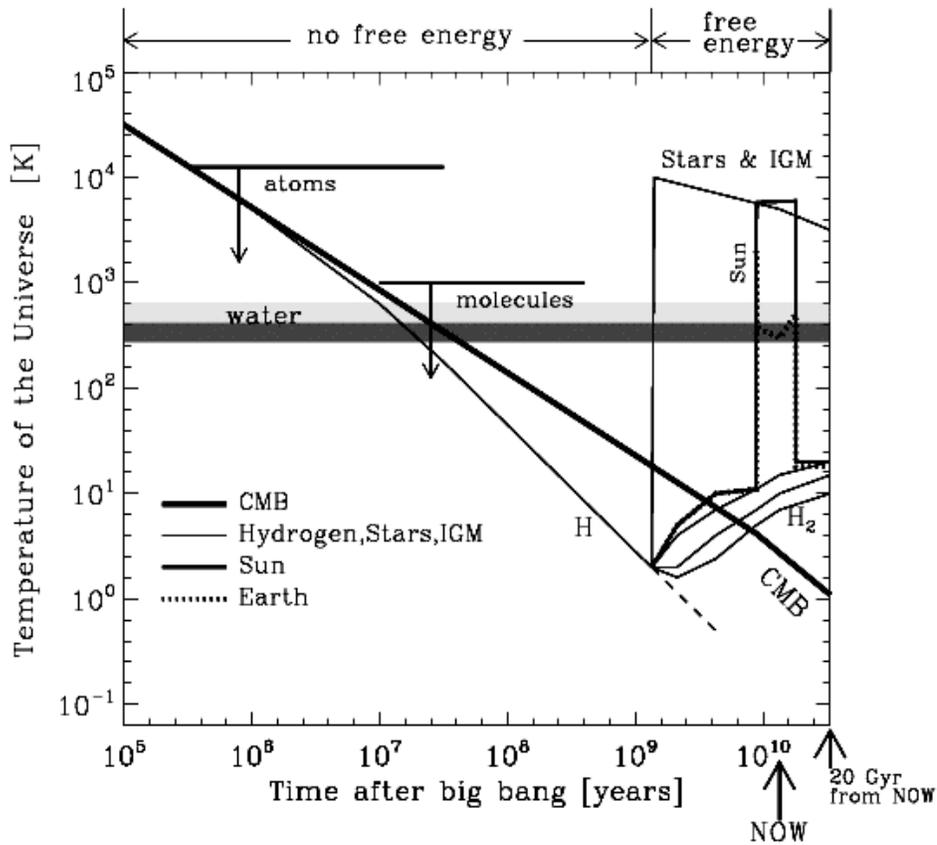

Figure 2. A zoom-in on the transitions in the lower right of Fig. 1. The thin line beneath the CMB line shows how hydrogen cooled more rapidly than the CMB. The dashed line shows how this cooling would have continued if it had not been for the fact that a billion years after the big bang, the thermal energy of the hydrogen sank low enough to allow the weak gravitational binding energy to contract the densest clouds of hydrogen. These clouds then became denser and hotter and eventually formed the first stars in the universe. These massive stars emitted UV photons that heated and ionized the more rarified hydrogen (intergalactic medium : IGM). The formation of these first massive stars and the re-ionization of the IGM are represented by the vertical line at $10^9$ years. The other lines originating at the same point represent the temperatures of hydrogen clouds that were dense enough to self-shield and avoid UV ionization. These clouds were gradually enriched with oxygen, carbon, nitrogen, iron and the other waste products of the supernovae explosions of the first, massive, short-lived stars. About 4.6 billion years ago one of these enriched clouds was shocked by a nearby supernova. This initiated collapse and star formation. One of the stars was the Sun. Planetary formation and formation of the Earth was part of this collapse (dotted line). Other, less dense clouds of $H_2$ (represented by the three other thin lines) collapsed a bit but stayed at 10 or 20 K. The upper x axis shows that free energy is available once the temperature of the hydrogen is low enough to initiate gravitational collapse and star formation. Two adjacent grey strips are labeled "water". The lower darker one is 0-100 C. The higher lighter one is 100-650 C; the highest temperatures at which water, under pressure, can exist.

Events on Mercury, Venus and Mars were similar: an initial period of heavy bombardment abated as planetesimals were swept up. The process of rocky planet formation in our Solar System and elsewhere involves many thermal constraints. The radially dependent temperature of the proto-stellar nebula segregates gases and solids. Stellar winds push the volatiles away, leave the refractories and fractionate elements and isotopes. Rocky planets start off hot, and cool with time. Smaller ones cool off more quickly. On larger ones, the gravitational energy of accretion and the decay of radioactive elements keep the planet hot and power convective currents in the mantle, widespread volcanism, active plate tectonics and large hydrothermal systems. These are probably the temperature constraints that any life in the universe has to begin with if it is to come into existence – these are universal features of hydrogen (enriched with ~1% of other elements) cooling in an expanding universe.

The hyperthermophilic monopoly on the deepest roots of the phylogenetic tree of terrestrial life (Fig. 3) suggests that temperature played a dominant role in biogenesis and/or dominated the selection pressure on the earliest forms of life. This suggests the more general idea that wherever biogenesis occurs in the universe, temperature will play a dominant role not only in setting the stage with stars, planets and the building blocks of life but also in determining, restricting and constraining the earliest and simplest forms of life. Hyperthermophiles may dominate not only the terrestrial tree of life but the trees of life of all planets in the universe. What we find out about terrestrial biogenesis may be representative of biogenesis throughout the universe.

Terrestrial life is made of H, O, C and N: the first, second, third and fourth most abundant elements in the universe (ignoring the chemically inert helium). If life exists elsewhere, pundits expect it to be based on carbon and water. Even if silicon were a possibility, carbon-based life would still be 10 times more abundant than silicon-based life simply because carbon is ten times more abundant in the universe than is silicon. $H_2O$ is the combination of the two most abundant chemically active elements in the universe. It is the most abundant tri-atomic molecule in the universe. $H_2O$ is everywhere. However liquid $H_2O$ is much less common and can only be found within a narrow range of temperatures (Fig. 2).

## 3. When Do the Sources of Free Energy in the Universe Permit Life to Exist?

In the beginning, 13.5 billion years ago, the universe was very hot (Lineweaver, 1999). There was no life and there were no structures in the universe. The universe was a thermal heat bath of photons. Life is not possible in a heat bath. In thermal equilibrium no free energy is available. Life will not emerge simply because the universe has become the right temperature for $H_2O$ to be a liquid. Life needs a source of free energy. The origin of all sources of free energy can be traced back to the expansion of the universe which led to the cooling of hydrogen, which led (one billion years after the big bang) to the formation of the first stars and planets.

The chemical composition also constrains life. Even with abundant free energy one billion years after the big bang, life could not have formed since water, life and Earth-like planets all require elements that did not exist in any abundance. In Lineweaver (2001) we showed that another billion years was required before Earth-like planets could form.

## 4. Phylogenetic Thermometry: Hot Ancestors and their Cool Descendants

By identifying the oldest features in extant life we can get an idea of what the thermal constraints on the earliest life were. Molecular biologists have been able to construct phylogenies based on immune responses, molecular weights, protein similarities and DNA and RNA sequences. To examine the deepest roots of the tree of life the most highly conserved homologous sequences are compared. The phylogenetic tree of life in Fig. 3 is based on 16S rRNA (Pace, 1997). Life can be usefully classified into three domains: Bacteria, Archaea and Eukarya (Woese et al., 1990). Trees based on other highly conserved DNA sequences are in good but not complete agreement with this tree. The position of the root of the tree is estimated from ancient gene duplications (Iwabe et al., 1989; Gogarten - Boekels et al., 1995; Grimaldo and Cammarano, 1998). Only autotrophic hyperthermophiles from hot springs and hydrothermal vents are found near the root. This suggests that the last common ancestors (LCA) lived in similarly hot environments (Shock et al., 2000; DiGiulio, 2000). The environment of the LCA is our best estimate for thermal constraints on the emergence of life.

TABLE 2.  Maximum Growth Termperatures  (see Fig. 3)

| Organism | Temperature [degrees C] | References |
|---|---|---|
| Archaea | | |
| *Methanopyrus* | 110 | Stetter, 1996 |
| *Pyrodictium* | 110 | Stetter, 1996 |
| *Thermoproteus* | 97 | Stetter, 1996 |
| *Methanothermus* | 97 | Stetter, 1996 |
| *Archaeoglobus* | 95 | Stetter, 1999; Brock et al., 1994 |
| *Thermofilum* | 95 | Stetter, 1996 |
| *Thermococcus* | 93 | Stetter, 1998 |
| *Methanococcus* | 91 | Stetter, 1999 |
| *Sulfolobus* | 87 | Brock et al., 1994 |
| *Methanobacterium* | 75 | Brock, 1978 |
| *Thermoplasma* | 67 | Kristjansson and Stetter, 1992 |
| *Haloferax* (Halophiles) | 55 | Kristjansson and Stetter, 1992 |
| Bacteria | | |
| *Aquifex* | 95 | Stetter, 1996 |
| *Thermotoga* | 90 | Stetter, 1996 |
| *Thermus* (*T. thermophilus*) | 85 | Kristjansson and Stetter, 1992 |
| *Bacillus* | 82 | Brock, 1978 |
| *Clostridium* | 78 | Wiegel, 1992 |
| *Synechoccoccus* | 74 | Brock, 1978 |
| *Chloroflexus* | 73 | Brock, 1978 |
| Mitochondrion/*Agrobacterium* | 60 | Brock et al., 1994 |
| *Desulfovibrio* | < 60 | Kristjansson, 1992 |
| *Heliobacterium* (*H. modesticaldum*) | 56 | Kimble et al., 1995 |
| *Chlorobium* | 55 | Sirevag, 1992 |
| *Flavobacterium* (*F. autothermophilum*) | 55 | Aragno, 1992 |
| *Plantomyces* (*Isophaera pallida*) | 55 | Schmidt and Starr, 1989 |
| Eukarya | < 60 | Madigan et al., 1997 |

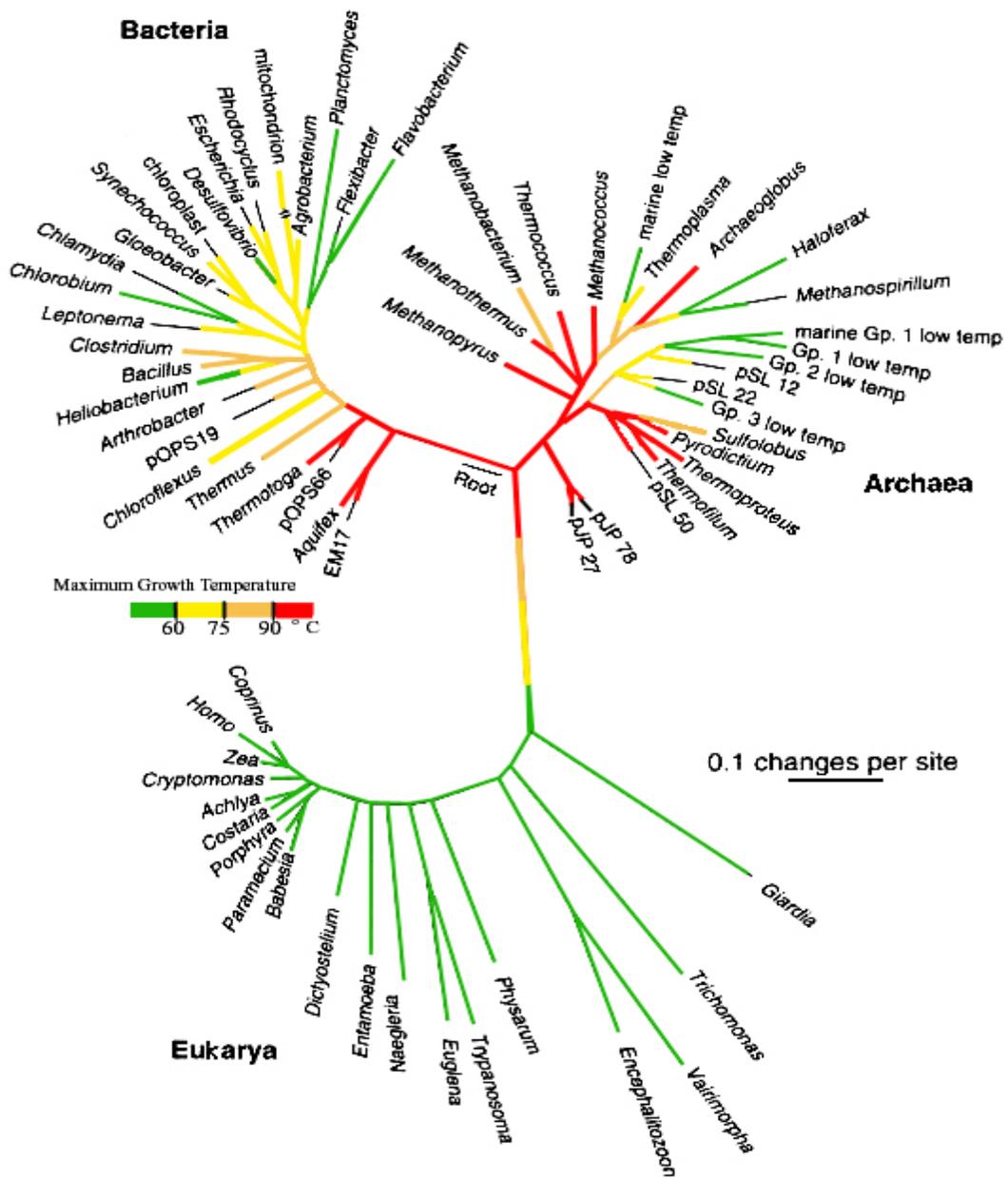

Figure 3. Phylogenetic tree of life based on 16S rRNA sequences. Maximal growth temperatures (Table 2) have been used to color-code the branches. See text and Pace (1997) for details.

Here we attempt to quantify the gradient between hot ancestors and cool descendants. The distance from the root to the tip of any branch is proportional to the number of changes per site (see scale bar). Hyperthermophiles are living fossils in the sense that they occupy the shortest branches – their 16S rRNA has undergone the fewest changes over the past ~ 4 billion years. The fact that hyperthermophiles have a monopoly on the root and that organisms that cannot tolerate heat are only found far from the root suggests that we can use this phylogenetic tree as a crude thermometer. In Schwartzman and Lineweaver (2003) and in Table 2 we assemble from the literature maximum growth temperatures ($T_{max}$) for as large a subsample of the species shown in Fig. 3, as possible. We have used these temperatures in Fig. 3 to color-code the tree (see temperature scale). Since values of $T_{max}$ are derived from extant organisms, the first step of the color coding was to color only the tips of the branches for which $T_{max}$ values were available. The tips of the branches for which we have no information are left as thin black lines.

Maximum growth temperature is determined by the fundamental biochemistry of the organism and is therefore a highly conserved feature. This motivates the plausible assumption that one could extend the color of the tip back through time at least to the closest node (a node is where two or more branches merge). When the branches from a node have the same color, we extend the same color one node closer to the root and so on. When the branches from a node do not share a color, we extend the coloring towards the root with the color of the branch that matches the next node. The end result is a phylogenetic thermometer. The self-consistency of this thermometer is not built in to the procedure to make it. For example, the phylogenetic position of *Archaeoglobus* among the Archaea is an inconsistency. Its $T_{max}$ is higher than the $T_{max}$ of two ancestral nodes closer to the root. If the $T_{max}$ of an organism were not a highly conserved trait, instead of yielding a highly consistent picture of high $T_{max}$ near the root and low $T_{max}$ further away, we would have a mosaic of colors with no identifiable pattern. There would be many more examples of *Archaeoglobus*-like inconsistency.

The simple picture that emerges from this analysis is of mesophiles evolving from hyperthermophiles and gradually adapting to cool environments. It is easily tested since it makes the simple prediction that branches ending in black tips (for which we have no information) will have $T_{max}$ values consistent with the colors closest to their tips (or slightly cooler). Exceptions like *Archaeoglobus*, for which this is not the case, will be rare or be indications of misplacement in the tree or an indication that *Archaeoglobus*' ability to tolerate heat is a recent adaptation, possibly due to substantial lateral transfer of genes responsible for heat tolerance. The most deeply rooted mesophile, "Gp. 3 low temp" (in the Archaea) is also a candidate for being misplaced in the tree. The few discrepancies between this 16S tree and other trees may correlate with these exceptions.

Figure 3 shows the Eukarya connected to the hyperthermophilic root with a transition from hyperthermophile to thermophile to mesophile. However, the distances from the root at which the ancestors of Eukarya became thermophiles and then mesophiles is unknown. Thus the positions of these transitions are guesses. An endosymbiotic origin for Eukarya even calls into question the idea that these transitions occurred along this branch (Martin and Russell, 2002). This phylogenetic thermometer would be completely undermined if it could be established that the ancestors of hyperthermophiles were mesophiles who evolved a heat tolerance as they learned to adapt to the hottest aqueous environments on Earth (Valley et al., 2002; Matte-Tailliez et al., 2002). A comparison of the antiquity of heat-shock proteins and 'cold-shock' proteins could test this idea; so might further comparative analyses of G + C content.

If life lived in cool environments 3 - 4 billion years ago, then we could expect mesophiles near the root. Either there were no such environments, or life did not live in them or they have left no survivors. A close examination of this idea's most plausible representative, "Gp. 3 low temp", is called for. Until more definitive tests are made, the simple interpretation of this phylogenetic tree, as a remarkably self-consistent thermometer based on a crude but quantifiable gradient in $T_{max}$, gives us one of the few methods we have for estimating the thermal history of life and the Earth over the past few billion years.

In Fig. 2 of Schwartzman and Lineweaver (2003) we plotted $T_{max}$ values as a function of branch length from the root to the nearest node of the four main trunks. A strong correlation was found. The self-consistent color gradient in Fig. 3 from hot root to cold branches is an alternative way to represent the same correlation.

**5. Time Calibration of the Phylogentic Thermometer**

Since thermometers on billion year time scales are so rare and approximate, it is important to compare the ones we do have. Our confidence in these thermometers will depend on their consistency. To turn this phylogenetic thermometer into a thermal record of the Earth's history we need to establish a connection between absolute time and the nodes in the main branches of the phylogenetic tree. We will consider the nodes along four main trunks each starting from the root and extending respectively to *Homo* in the Eukarya, *Planctomyces* in the Bacteria, *Methanospirillum* in the Euryarchaeota and "marine Gp. 1 low temp" in Korarchaeota/Crenarchaeota. Along each of these main trunks are nodes. The closer the nodes are to the root, the earlier the time associated with that node. This is a widely accepted interpretation of the relative chronology of these nodes. To convert these relative chronologies into absolute chronologies we will make several plausible assumptions. The end result will be a plausible, approximate time calibration of these nodes with the virtue of being explicit and therefore open for improvement.

First we assume that the LCA lived 4 billion years ago. This is based on the phenotypic similarity of 3.5 Gyr old fossils to extant organisms (e.g. Schopf, 1994) with a small correction for incompleteness (for an analysis of correction factors applied to an incomplete fossil record see Martin, 1993). Thus, we are assuming that the distance from the root to any tip corresponds to the same duration: 4 Gyr. This gives us a measure of how variable the molecular clocks in the various branches are since the dispersion in the distances from the root to the tips of the tree is a measure of the dispersion in the speeds of the molecular clocks.

We also adopt several absolute calibration points. For example, much work has gone into establishing the time of the most recent common ancestor of animals, plants and fungi (node in Eukarya of Fig. 3 where *Homo*, *Zea* and *Coprinus* merge). We adopt 1.6 billion years ago as the time of this node (Hedges, 2002). We then have the constraint that the distance from this node to the root corresponds to 2.4 Gyr (= 4.0 – 1.6). We then make a simple linear mapping between the nodes along this distance and the 2.4 Gyr time interval. Although this mapping has the virtue of simplicity, Woese and others have pointed out a general trend for molecular clocks to slow down with age. We have considered several non-linear scalings to correct for this trend. These corrections compress the horizontal distances between the points on the left side of Fig. 4 but do not change the main result. We have also used values between 2.4 and 2.8 Gyr for the emergence of cyanobacteria (*Synechococcus*/chloroplast node). In this way we obtain a plausible but rough time calibration of Fig. 3 using branch lengths from the root to various nodes along the four main trunks mentioned above. The results from these four trunks are plotted in Fig. 4.

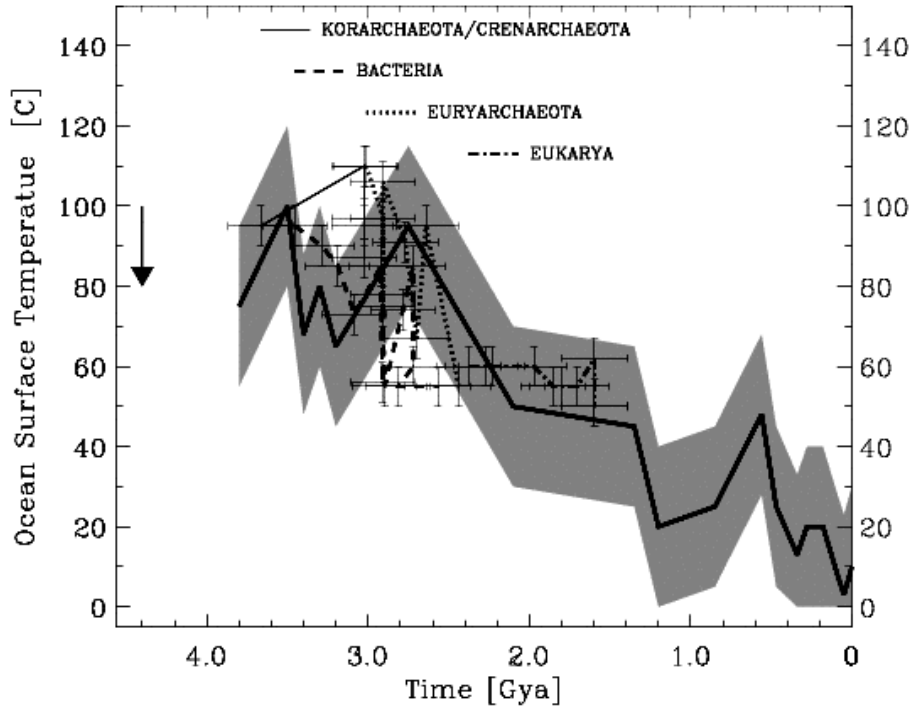

Figure 4. Temperature history of the surface of the Earth based on $^{18}O$ chert values. The thick solid line is a slightly smoothed version of data from (Knauth, 1992 and Knauth and Lowe, 2003). The grey band is an error estimate. Superimposed on this are the estimated times and $T_{max}$ values of the nodes along the four main trunks of the tree in Fig. 3 (see text).

## 6. Comparison of the Phylogenetic and $^{18}$O Chert Thermometers: Implications for the Origin of Life.

In Fig. 4 we compare the phylogenetic thermometer of the temperature of the Earth over the past three to four billion years to the $^{18}$O chert thermometer (Knauth, 1992; Knauth and Lowe, 2003). According to the chert data, over the last four billion years ocean surface water has cooled from ~90 C to ~10 C. From 3.5 Gyr ago to 1.5 Gyr ago it decreased from 80 C to 50 C. Over this same time interval, the $T_{max}$ of newly evolved organisms decreased in approximately the same way.

Fig. 3 shows clearly that the roots of the tree of life are hyperthermophilic. Fig. 4 shows an easily identifiable correlation between decreasing temperatures based on $^{18}$O in chert and the decreasing $T_{max}$ values of newly evolving organisms at the nodes of the four main trunks in Fig. 3. These two thermometers seem to be consistent. Their similarity helps explain why mesophiles seem to have thermophile ancestors but not vice versa. If we take the $^{18}$O chert thermal history seriously, 4 billion years ago there were no environments for mesophiles to live in. If this is true then deep molecular phylogenies of thermophiles, mesophiles and psychrophiles will continue to show a pattern of ancestral thermophily with mesophily and psychrophily as derived adaptation. The evolution from hot to cold could be interpreted as hot environments being more difficult to adapt to than cold ones; an asymmetry in adaptive ability. However, the thermal history from $^{18}$O chert suggests that there were only hot environments and these were already occupied. The new challenge for life was the adaptation to cooler places as they emerged. This seems to be the most straight-forward reading of the good but rough correlation between the phylogenetic and $^{18}$O chert thermometers. However, although the absolute dates of the $^{18}$O thermometer are well established, their interpretation as a measure of ocean surface water temperature is controversial (Des Marais, Kasting, private communication).

We have strong evidence on Earth that life started out simple and that the simplest, most basic pieces of our metabolic pathways are the most ancient. We assume that if there is life on other planets that it too started out simple. Therefore the best candidates for universal features of life are the most ancient features of life that we are able to identify here on Earth. We should expect the most fundamental pathways close to life's origin to have been explored by life elsewhere.

The deepest rooted divergence in the tree of life in Fig. 3 is between the Archaea and the Bacteria. Although we have no examples of extraterrestrial life, we believe it is meaningful to speculate about whether a similar divergence occurs universally in extraterrestrial trees of life. Psychrophiles (cold loving bacteria) have membrane lipids rich in unsaturated fatty acids making the membranes more fluid and flexible at low temperatures. Thermophiles have membrane lipids rich in saturated fatty acids giving them stability at high temperatures since saturated fatty acids form much stronger hydrophobic bonds (Madigan et al., 1997). Hyperthermophiles, virtually all of which are Archaea, do not contain fatty acids in the lipids of their membranes but instead have hydrocarbons of various lengths; phytane with ether linkages (rather than ester linkages in Bacteria). Are these the only two possibilities for membranes to emerge in high temperature environments? This ether/ester, temperature-related structural difference between Archaea and Bacteria is a good candidate for a universal divergence – a divergence that may be quasi-deterministic – a step in the emergence of life that may be less deterministic than the features we have been discussing but more deterministic than downstream contingencies like the eukaryotic divergence ~2.5 billion years later between plants, animals and fungi.

## 7. First it was Hot, then it got Cool

The transition from hot to cool happens to cups of coffee. It happens to planets and it happens to life (Figs. 3 and 4). It is either the case that life emerges as soon as it can (in which case it will emerge when the environment is still at the hot end compatible with liquid water) or the emergence of life depends on a hot environment (in which case it will emerge *only* while the environment is at the hot end compatible with liquid water).

The connections between temperature and life are so fundamental that we can use them as guides in our estimates of what life forms we can reasonably expect to exist beyond the Earth. Hyperthermophiles may dominate not only the terrestrial tree of life but the trees of life of all planets in the universe.

Extant life is no longer a passive inhabitant of a given ambient temperature. The relation between life and temperature has become a quasi-deterministic one in the sense that although temperature has played a dominant role in constraining life, life has also been able to modify the temperature within limits set by the deterministic processes of planet formation and the evolution of stellar luminosity (Schwartzman 1999). The nature of this temperature regulation is a hotly debated central conjecture of the Gaia hypothesis (Schneider and Boston, 1991).

It may be the case that we are oblivious to the thermal straight-jacket that our biochemistry has to wear. The apparent quirkiness of biological evolution may only reflect our obsession with the changing fashions of phenotype rather than our biochemical foundations. Maybe the evolution of biochemistry today is just as thermally constrained and deterministic as it was 4 billion years ago. Maybe that is why our metabolic paths are so ancient.